\documentclass[aps,twocolumn,nofootinbib, preprintnumbers, superscriptaddress]{revtex4}

\usepackage{amsmath, amssymb, slashed, braket, bm}
\usepackage{graphicx}
\usepackage{epstopdf}
\usepackage{float,appendix}
\usepackage[colorlinks=true,
            linkcolor=blue,
            urlcolor=blue,
            citecolor=green,          
            bookmarks=true,
            bookmarksnumbered=true,
            breaklinks=true,
            pdfpagemode=FullScreen,
            pdfstartview=FitBH]{hyperref}
\usepackage{esint}
\usepackage[normalem]{ulem}
\usepackage{siunitx}
\usepackage{multirow}
\usepackage{xcolor}

\graphicspath{{figs/}} 

\usepackage{xcolor}
\definecolor{gesfpurple}{rgb}{0.47,0.19,0.42}

\definecolor{gesflanse}{rgb}{0.00,0.50,0.50}

\definecolor{gesfblue}{rgb}{0.08,0.42,0.76}

\definecolor{gesfred}{rgb}{1,0,0}

\definecolor{gesfwhite}{rgb}{1,1,1}

\definecolor{gesfblack}{rgb}{0,0,0}

\newcommand{\gapp}[1]{{\hypersetup{linkcolor=red}App.\,\ref{#1}\hypersetup{linkcolor=blue}}}
\newcommand{\geqn}[1]{Eq.\,\hypersetup{linkcolor=blue}(\ref{#1})\hypersetup{linkcolor=blue}}
\newcommand{\gfig}[1]{{\hypersetup{linkcolor=violet}Fig.\,\ref{#1}\hypersetup{linkcolor=blue}}}

\newcommand{\prlsection}[2]{{\it\textbf{#1}{#2}}---}

\begin{document}

\title{
Enhanced Rydberg-Atom Superheterodyne Detection \\ of Hidden-Photon Dark Matter on Chips
}

\author{Xiaochen Li}
\affiliation{State Key Laboratory of Micro-nano Engineering Science, Tsung-Dao Lee 
Institute, Shanghai Jiao Tong University, Shanghai 201210, China}

\author{Bo Gao}
\email{gaobo$\_$79@sjtu.edu.cn}
\affiliation{State Key Laboratory of Micro-nano Engineering Science, Tsung-Dao Lee Institute, Shanghai Jiao Tong University, Shanghai 201210, China}
\affiliation{Shanghai Research Center for Quantum Sciences, Shanghai 201315, China}

\author{Shigeki Matsumoto}
\email{shigeki.matsumoto@ipmu.jp}
\affiliation{Kavli IPMU (WPI), UTIAS, University of Tokyo, Kashiwa, 277-8583, Japan}

\author{Jie Sheng}
\email[Corresponding author: ]{jie.sheng@ipmu.jp}
\affiliation{Kavli IPMU (WPI), UTIAS, University of Tokyo, Kashiwa, 277-8583, Japan}

\author{Chuan-Yang Xing}
\email{cyxing@upc.edu.cn}
\affiliation{College of Science, China University of Petroleum (East China), Qingdao 266580, China}

\author{Hong Ding}
\affiliation{State Key Laboratory of Micro-nano Engineering Science, Tsung-Dao Lee 
Institute, Shanghai Jiao Tong University, Shanghai 201210, China}

\begin{abstract}

\noindent
Although hidden-photon dark matter with masses above $10^{-4}\,\mathrm{eV}$ is well motivated by inflationary production, it remains largely unexplored by terrestrial experiments. Through kinetic mixing, hidden photons induce a weak oscillating electric field above $10\,\mathrm{GHz}$. We propose to amplify this signal using a compact high-frequency distributed cavity and detect it with chip-scale Rydberg-atom superheterodyne spectroscopy. Combining resonant enhancement, large dipole moments of Rydberg atoms, and long-term stable integration, this approach can probe hidden-photon dark matter in the mass range $5 \times 10^{-5}\text{--}7\times 10^{-4}\,\mathrm{eV}$ with sensitivities $3$--$4$ orders of magnitude beyond existing limits.
\end{abstract}

\maketitle 

\noindent
\prlsection{Introduction}{.}
Many astrophysical and cosmological observations indicate that dark matter (DM) constitutes approximately $85\%$ of the total matter content of the Universe~\cite{Planck:2018vyg, Cirelli:2024ssz}, yet its particle nature remains one of the central mysteries in modern physics~\cite{Bertone:2004pz, Bertone:2016nfn}. Among the many candidates, the ultralight hidden-photon stands out as one of the most attractive possibilities~\cite{Arza:2026rsl}. As the gauge boson of an additional $U(1)$ symmetry~\cite{Holdom:1985ag, Arias:2012az, Fabbrichesi:2020wbt}, it can couple to the Standard Model (SM) photon through kinetic mixing. It thus provides one of the simplest extensions of the SM, while exhibiting a rich phenomenology closely analogous to that of electromagnetism~\cite{Caputo:2021eaa}.

From a theoretical perspective, the meV scale is a particularly important and well-motivated mass range for hidden-photon DM. If such a gauge boson exists, it can be produced naturally during inflation through quantum fluctuations~\cite{Graham:2015rva}. In this case, the present-day abundance is determined largely by the hidden-photon mass, $m_{A'}$, and the Hubble scale during inflation, $H_I$, and this abundance matches the observed DM abundance when~\cite{Graham:2015rva}
\begin{equation}
    \Omega_{A'} \simeq 0.3 \sqrt{\frac{m_{A'}}{10^{-3}\,\text{eV}}} \left( \frac{H_I}{3 \times 10^{13}\,\text{GeV}}\right)^2.
    \label{relic}
\end{equation}
Therefore, high-scale inflation with $H_I \sim 10^{13}~\mathrm{GeV}$ naturally points to hidden-photon DM in the meV mass range. Plateau-like inflation models with a large Hubble scale are favored by current CMB data~\cite{Starobinsky:1980te, Bezrukov:2007ep, Kallosh:2013yoa}. Current limits on primordial tensor modes imply $H_I \lesssim 4.7\times 10^{13}~\mathrm{GeV}$~\cite{Planck:2018vyg, BICEPKeck:2021gln}, which in turn yields a lower bound on the hidden-photon mass, $m_{A'} \gtrsim 1.5\times 10^{-4}~\mathrm{eV}$.

Experimentally, the search for such light hidden-photon fields and an additional $U(1)$ symmetry has driven the development of highly sensitive detectors and quantum-metrology techniques~\cite{Jaeckel:2007ch,Arias:2012az,Horns:2012jf,Chaudhuri:2014dla,Arias:2014ela,Chaudhuri:2018rqn,Chen:2022quj,An:2022hhb,PhysRevD.108.035042,Cheng:2024yrn,PhysRevD.109.032009,Gao:2025ryi,Chigusa:2025rqs,Zheng:2025qgv,banerjee2025rydbergsinglephotondetection,Ma:2026ujz,Chen:2026yss,Xing:2026pyo,Davoudiasl:2026opo}. Among these efforts, haloscope experiments based on resonant electromagnetic structures have achieved leading sensitivity in searches for hidden-photon DM~\cite{PhysRevLett.51.1415,PhysRevD.90.075017,Arias:2014ela,PhysRevLett.120.151301,PhysRevLett.124.101303,CAPP:2024axion,HAYSTAC:2021squeezing,HAYSTAC:2024phaseII,ORGAN:2024,QUAX:2024,MADMAX:2024first,CAST-RADES:2021,DarkSRF:2023,SHANHE:2024}. However, their reach deteriorates rapidly at high frequencies, where the resonator volume decreases with the wavelength, leading to a strong suppression of the signal power~\cite{Aybas:2026rwu}. This limitation is particularly severe in the meV mass range. It is therefore crucial to develop new detection mechanisms targeting this mass range.

Rydberg atoms, whose valence electrons occupy highly excited states, possess large electric dipole moments, while the energy spacing between neighboring Rydberg levels naturally lies in the meV range. This makes them a powerful platform for precision microwave electric-field sensing~\cite{Liu:2023zxe}. Meanwhile, rapid progress in neutral-atom quantum technologies has greatly advanced the experimental toolbox for controlling and probing Rydberg systems~\cite{2010QuantuminformationRydbergatoms,Liu:2023zxe,PhysRevD.109.032009,PhysRevResearch.6.023017,Barik:2024emn}. In this \textit{Letter}, we propose, for the first time, that the high-frequency weak electric field induced by hidden-photon DM can be probed using a Rydberg-atom superheterodyne detection scheme. The rich level structure of Rydberg states is naturally suited to signals in the $\mathcal{O}(10)\,\mathrm{GHz}$ regime. Combining long integration times enabled by recently developed chip-scale atomic vapor cells with signal enhancement from a compact cavity, this strategy can achieve improved sensitivity to hidden-photon couplings in the meV mass range. Throughout this Letter, we use natural units with $c=\hbar=1$.

\vspace{0.35cm}

\noindent
\prlsection{Cavity-Enhanced Hidden-Photon Field}{.}
We consider the minimal hidden-photon DM model, in which a hidden $U(1)'$ gauge boson $A'_\mu$ couples to the SM photon through kinetic mixing $\epsilon$. The Lagrangian is given by
\begin{equation}
    \mathcal{L} =
    -\frac{1}{4}F'_{\mu\nu}F'^{\mu\nu}
    +\frac{1}{2}m_{A'}^{2}A'_{\mu}A'^{\mu}
    -\frac{1}{4}\epsilon F'_{\mu\nu}F^{\mu\nu}
    +\mathcal{L}_{\rm SM},
\end{equation}
where $F'_{\mu\nu}$ and $F_{\mu\nu}$ are the hidden- and ordinary-photon field strengths, and $m_{A'}$ is the hidden-photon mass generated through the Stueckelberg mechanism~\cite{Ruegg:2003ps}.

With its abundance generated by inflationary quantum fluctuations, as summarized in \geqn{relic}, hidden-photon cold DM can be treated today on laboratory scales as non-relativistic massive Proca modes. Each mode induces an oscillating dark electric field, whose total amplitude follows by summing all contributions\,\cite{Arias:2012az, Fabbrichesi:2020wbt}
\begin{eqnarray}
    {\bf E}'
    &=& \sum_i {\bf E}'_i \cos(E_i t - {\bf p}_i \cdot {\bf x}_i + \phi_i)
    \nonumber \\
    &\simeq&
    \sqrt{2\rho_{\rm DM}}\,{\bf e} \cos(m_{A'}t + \phi),
\end{eqnarray}
where $i$ labels each DM contribution, characterized by the induced dark electric field ${\bf E}'_i$, energy $E_i \simeq m_{A'}$, momentum ${\bf p}_i$, and phase $\phi_i$. Since Galactic DM has velocity dispersion $v \sim 10^{-3}$, each hidden-photon DM mode carries momentum $|{\bf p}_i| \sim m_{A'}v \sim 10^{-3}m_{A'}$, leading to a kinetic-energy spread and hence a fractional frequency bandwidth $\Delta\omega/\omega \sim v^2 \sim 10^{-6}$. Therefore, the final form of the expression above is valid only within the coherence time $\tau_c \simeq 1/(m_{A'}v^2) \simeq 7\,(10^{-4}\,{\rm eV}/m_{A'})\,\mu{\rm s}$ and coherence length $\ell_c \simeq 1/(m_{A'}v) \simeq 2\,(10^{-4}\,{\rm eV}/m_{A'})\,{\rm m}$, where ${\bf e}$ denotes the polarization vector. Here, $\phi$ is the effective DM-field phase; it is approximately constant within a coherence patch of size $\tau_c$ and $\ell_c$, but becomes randomized between patches. The normalization of the dark electric field is fixed by the local DM density. Taking $\rho_{\rm DM}=0.45\,{\rm GeV}/{\rm cm}^3$, one obtains $\sqrt{2\rho_{\rm DM}} \simeq 4\,{\rm kV/m}$.

The oscillating dark electric field generates an extremely weak ordinary oscillating electric field with wavelength $\lambda \sim 2\pi/m_{A'}$ through kinetic-mixing-induced interactions with materials~\cite{Caputo:2021eaa}. In the meV hidden-photon mass range of interest, current constraints on the kinetic mixing parameter have reached the level of $\epsilon \sim 10^{-9}$~\cite{Arias:2012az}. Accordingly, our target signal is an ultra-weak oscillating electric field with an amplitude $\lesssim \epsilon\sqrt{2\rho_{\rm DM}} \simeq 4.0 \times 10^{-6}\,\mathrm{V/m}$ and a frequency $m_{A'} \gtrsim \mathcal{O}(10\,\mathrm{GHz})$. 

To enhance this hidden-photon-induced field to a detectable level, we employ a microwave distributed cavity. When the DM frequency is resonant with a cavity eigenmode, it coherently drives the corresponding mode and enhances the electric field into an observable signal:
\begin{equation}
    E_s =
    4.0 \times 10^{-6}\,\frac{\text{V}}{\text{m}} \times \left(\frac{\epsilon}{10^{-9}}\right) \times A,
    \label{eq:Etarget}
\end{equation} 
by an amplification factor $A \equiv Q|\eta_{\rm field}|$, which is determined by the loaded quality factor $Q$ and a mode-dependent field-response factor $\eta_{\rm field}$. The latter quantifies how well the cavity electric-field profile matches the direction and distribution of the DM-induced drive~\cite{sikivie1983experimental}. Details of the cavity-enhanced electric field appear in App.~\ref{cavityMechanism}. For the cylindrical cavity discussed later, the resonant frequency of the $\mathrm{TM}_{mnp}$ mode is given by
\begin{equation}
    f^{\mathrm{TM}}_{mnp} =
    \frac{1}{2\pi}
    \sqrt{
    \left(\frac{\chi_{mn}}{R}\right)^2
    +\left(\frac{p\pi}{L}\right)^2},
    \label{eq:tm0n0}
\end{equation}
where $R$ and $L$ are the radius and length of the cavity, respectively, and $\chi_{mn}$ denotes the $n$-th radial zero of the Bessel function $J_m$~\cite{balanis2012advanced}. Taking the lowest mode, $\mathrm{TM}_{010}$, as an illustrative example, frequencies above $\mathcal{O}(100)\,\mathrm{GHz}$ correspond to cavity radii around the $\mathcal{O}(\mathrm{mm})$ scale.

Conventional hidden-photon cavity searches become challenging at high frequencies: the cavity volume shrinks as the frequency increases, while the signal power scales with the effective volume~\cite{Arias:2012az}. As discussed below, by using chip-scale vapor cells compatible with small high-frequency cavities, we employ Rydberg atoms to directly and effectively probe the cavity-enhanced electric field, thereby achieving a greater sensitivity gain in $\epsilon$.

\vspace{0.35cm}

\noindent
\prlsection{Rydberg-Atom Superheterodyne Spectroscopy for High-Frequency Electric-Field Detection}{.}
Rydberg atoms provide a natural platform for detecting weak microwave electric fields. For a field with amplitude $E_s$ and frequency $\omega$ near a selected Rydberg transition, the signal induces an effective coupling between neighboring Rydberg levels, characterized by the Rabi frequency
\begin{equation}
    \Omega_s = \mu E_s
    \label{eq:Omega_s_main}
\end{equation}
where $\mu$ is the large transition dipole matrix element.

First, the energy spacing between neighboring Rydberg states naturally lies in the relevant range, $10^{-4}$--$10^{-3}\,\mathrm{eV}$. For a Rydberg electron in the state $|nlj\rangle$, the binding energy is well described by the Rydberg--Ritz formula, $E_{nlj}=-R/(n^\ast)^2$, where $R\simeq 13.6\,\mathrm{eV}$ is the Rydberg constant and $n^\ast$ is the effective principal quantum number including the quantum-defect correction~\cite{Gallagher1994RydbergAtoms}. In typical experiments, the two Rydberg levels involved differ in principal quantum number by $\Delta n=0$ or $1$, while $\Delta n^\ast$ remains of order unity even for states with different orbital angular momenta. The transition energy therefore follows the well-known $n^{-3}$ scaling~\cite{Gallagher1994RydbergAtoms},
\begin{equation}
    \Delta E \equiv
    E_{n^\ast+\Delta n^\ast}-E_{n^\ast}
    \approx \frac{2R\,\Delta n^\ast}{(n^\ast)^3} \approx \frac{2R}{(n^\ast)^3}.
    \label{DeltaE}
\end{equation}
For experimentally relevant Rydberg states with $n^\ast \sim 20$--$80$~\cite{song2018field}, the corresponding transition energies span
\begin{equation}
    \Delta E =  
    5.3\times10^{-5}\text{--}3.4\times10^{-3}\,\mathrm{eV}.
    \label{levelbounds}
\end{equation}
This makes the detection scheme naturally sensitive to
sub-meV hidden-photon DM at all accessible Rydberg-transition points within this mass window\footnote{
    To scan continuously across this mass range in practical implementations, one can further exploit magnetic-field tuning of the selected Rydberg transition, including both the linear Zeeman shift and, when relevant, the quadratic diamagnetic correction~\cite{Chigusa:2025rqs}; See \gapp{levels}.}. 

As noted above, Rydberg atoms are highly sensitive to electric fields owing to large electric dipole matrix elements, scaling as $\mu \sim n^2$~\cite{Liu:2023zxe}. Nevertheless, the DM-induced electric field \geqn{eq:Etarget}, and hence the coupling $\Omega_s$, remains too weak to be directly resolved with resonant cavity enhancement~\cite{zhou2025high}. We point out, however, that a Rydberg-atom superheterodyne scheme~\cite{jing2020atomic} can convert this weak high-frequency microwave coupling efficiently into an experimentally accessible signal.

\begin{figure*}[htp!]
    \centering
    \includegraphics[width=0.95\linewidth]{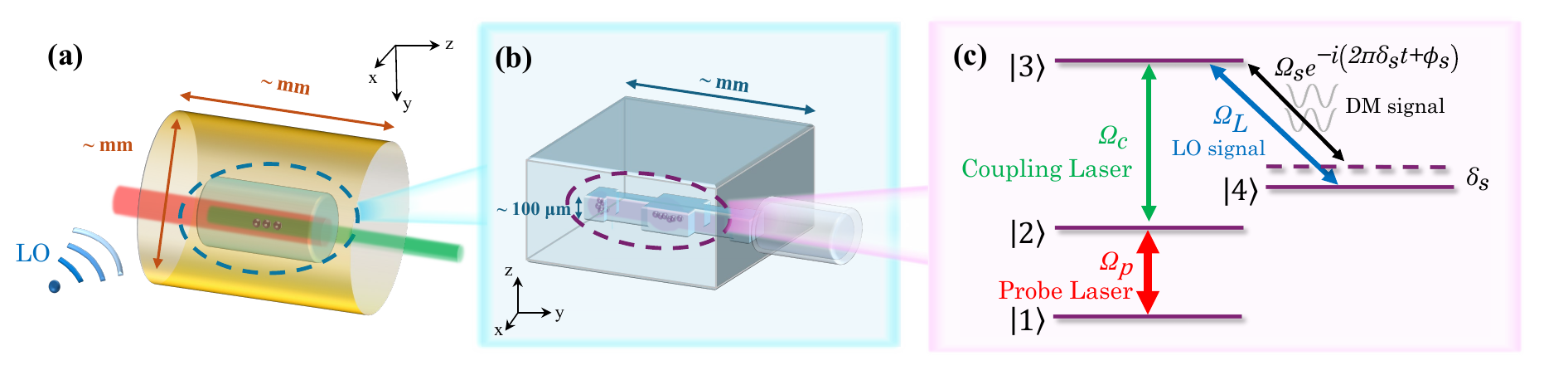}
    \caption{Schematic illustration of hidden-photon DM detection using a Rydberg-atom superheterodyne scheme.
    (a) Proposed experimental setup with an atomic vapor cell placed inside a resonant microwave cavity.
    (b) Chip-scale Rydberg-atom vapor cell used for optical readout.
    (c) Response of the superheterodyne detection scheme to the DM-induced electric field.}
    \label{fig:setup}
\end{figure*}

The superheterodyne protocol is based on ladder electromagnetically induced transparency (EIT)~\cite{fleischhauer2005electromagnetically,mohapatra2007coherent}, which provides an optical readout of the atomic response. The relevant EIT subsystem, shown in \gfig{fig:setup}(c), comprises the ground state $|1\rangle$, intermediate state $|2\rangle$, and Rydberg state $|3\rangle$. A weak probe laser drives the $|1\rangle \leftrightarrow |2\rangle$ transition, while a strong coupling laser drives $|2\rangle \leftrightarrow |3\rangle$. Under two-photon resonance~\cite{fleischhauer2005electromagnetically}, destructive interference cancels excitation to the intermediate state $|2\rangle$, trapping the atom in the corresponding dark state,
\begin{equation}
    |D\rangle =
    \frac{\Omega_c |1\rangle - \Omega_p |3\rangle}{\sqrt{\Omega_c^2+\Omega_p^2}},
    \label{eq:darkstate_main}
\end{equation}
where $\Omega_p$ and $\Omega_c$ denote the Rabi frequencies of the probe and coupling lasers, respectively. This produces a narrow transparency window in the probe transmission.

A strong local-oscillator (LO) microwave field is then applied to drive the transition between $\ket{3}$ and a nearby Rydberg state $\ket{4}$. It dresses the two Rydberg levels and splits the original dark state, and hence the EIT resonance, into an Autler--Townes doublet~\cite{autler1955stark}. If a much weaker DM-induced signal field is present on the transition, with $\Omega_s \ll \Omega_L$, the total microwave coupling is
\begin{equation}
    \Omega_{\rm tot}(t) \simeq \Omega_L + \Omega_s \cos(2\pi \delta_s t + \phi_s),
    \label{eq:Omega_tot_main_v2}
\end{equation}
where $\delta_s \equiv |m_{A'}-\omega_L|/(2\pi)$ is the beat frequency between the DM signal and the LO, $\omega_L$ is the angular frequency of the LO microwave field, and $\phi_s$ is their relative phase. The weak field therefore need not produce a resolvable spectral splitting on its own; instead, it modulates the LO-dressed Rydberg spectrum as a perturbation.

By setting the optical working point near the steep slope of the split EIT feature, even a small shift of the dressed EIT resonance is linearly converted into a measurable modulation of the transmitted probe power,
\begin{equation}
    \delta P(t)=\kappa \Omega_s \cos(2\pi \delta_s t+\phi_s),
    \label{eq:dP_main_v2}
\end{equation}
where $\kappa$ is the conversion coefficient determined by the local slope of the dressed EIT spectrum. In this way, the hidden-photon-induced high-frequency electric field is coherently mixed with the LO and down-converted into a low-frequency beat note, while preserving its amplitude and phase information. Since $\delta_s < 1\,$MHz~\cite{schlossberger2026fundamental}, efficient detection works when the signal frequency nearly matches the transition frequency between $\ket{3}$ and $\ket{4}$.

\vspace{0.35cm}

\noindent
\prlsection{Experimental Setup}{.}
A possible implementation of the scheme is shown in \gfig{fig:setup}. The apparatus combines a chip-scale cesium vapor cell~\cite{xing2025chip} for Rydberg-atom superheterodyne readout with a compact cavity for resonant enhancement of the hidden-photon-induced field.

\vspace{0.15cm}

\noindent
{\bf Chip-Scale Atomic Vapor Cell:}
We employ a mature chip-scale cesium vapor cell as the atomic sensor in the proposed experiment, as shown in \gfig{fig:setup}(b). This microfabricated cesium cell provides a well-established platform for Rydberg-atom superheterodyne readout, offering stable optical access and enabling long integration times of order $T \sim \mathcal{O}(1000)\,$s~\cite{wang2026sensing}. Recent experiments have reported integration times as long as $5000\,$s~\cite{jing2020atomic}.

The vapor cell employs a three-layer fused-silica sandwich structure, in which the internal cavity is defined in the middle substrate and sealed by two cover plates. The middle fused-silica layer has a thickness of $1\mathrm{mm}$~\cite{xing2025chip}, and its internal geometry is fabricated by femtosecond-laser micromachining followed by wet chemical etching. We assume a vapor cell size of $(2.5\mathrm{mm})^3$. The resulting microstructured cesium cell defines an atomic sensing region with a typical size of $\mathcal{O}(100)\mu\mathrm{m}$~\cite{xing2025chip}, so that the microwave electric field is sampled only within a highly localized volume. This more effectively suppresses unwanted spatial averaging of the signal and helps maintain an approximately uniform field over the atomic ensemble.

Cesium vapor is introduced by vacuum filling, with excess metallic Cs stored in an attached reservoir away from the main interaction region. During near-room-temperature operation, this reservoir stabilizes the vapor density and enables long-term optical readout~\cite{xing2025chip}.

The optical readout is implemented using a four-level ladder scheme in cesium, consisting of $|1\rangle = |6S_{1/2}, F=4\rangle$, $|2\rangle = |6P_{3/2}, F=5\rangle$, $|3\rangle = |nD_{5/2}\rangle$, and $|4\rangle = |(n+1)P_{3/2}\rangle$ or $|nF_{7/2}\rangle$~\cite{yuan2023quantum}.  A weak probe laser at $852\,\mathrm{nm}$ drives the $|1\rangle \leftrightarrow |2\rangle$ transition, while a coupling laser at $510\,\mathrm{nm}$ excites the $|2\rangle \leftrightarrow |3\rangle$ transition~\cite{fan2015effect}. The states $|3\rangle$ and $|4\rangle$ can be tuned to target different frequencies. The transmitted probe is continuously monitored using a balanced differential detection scheme, which suppresses common-mode laser noise and thereby improves the signal-to-noise ratio of the Rydberg-EIT readout.

\vspace{0.15cm}

\noindent
{\bf Distributed Cavity:} The prepared atomic vapor cell is placed inside the cavity~\cite{zhou2025high}, as shown in \gfig{fig:setup}(a), to sense the DM signal after resonant enhancement. Another key feature of this work is that the compact footprint of the chip-scale atomic vapor sensor is naturally compatible with the small spatial scale of the high-frequency distributed cavity used in this setup.

The minimum cavity size is determined by the dimensions of the atomic vapor cell. For a chip-scale cell, the smallest experimentally demonstrated thickness is approximately $2.5\,\mathrm{mm}$~\cite{xing2025chip}. We therefore consider a minimal cylindrical cavity with radius $R\simeq 2.5\,\mathrm{mm}$ and length $L\simeq 3\,\mathrm{mm}$. Small optical access apertures are introduced at both cavity end caps to allow lasers to pass through, while preserving an otherwise closed microwave resonator geometry. The LO microwave field is injected into the open cylindrical cavity through the apertures or weak side-coupling ports. 
Inside the cavity, it overlaps spatially with the cavity-enhanced signal mode and interacts with the Rydberg atoms, enabling atomic heterodyne mixing. The beat signal is then generated by the interference between the intracavity signal field and the injected LO field.

The corresponding fundamental frequency of the TM$_{010}$ mode in this cavity is $f^{\rm TM}_{010}=45.9\,\mathrm{GHz}$. Additionally, the use of higher-order modes is justified by the small active sensing volume of the Rydberg atoms. Although higher radial TM modes contain additional field nodes, the atomic interrogation region in our setup is $\mathcal{O}(100)~\mu\mathrm{m}$, much smaller than the cavity radius, $R=2.5~\mathrm{mm}$. Therefore, over the sub-millimeter atomic sensing region centered at $r=0$, the field remains close to its maximum even for the TM$_{020}$ and TM$_{030}$ modes. Consequently, these modes can be used to access an even higher resonant frequency range. The highest frequency considered here is therefore set by the TM$_{030}$ mode,
\begin{equation}
    f_{\rm max} =  f^{\rm TM}_{030}=165.2~\mathrm{GHz}.
    \label{eq:mmax_cavity}
\end{equation}
For example, the atomic transition between 
$\ket{31 D_{5/2}}$ and $\ket{28 F_{7/2}}$ matches such a frequency. This transition frequency can be converted into an upper bound on the detectable DM mass through the relation $f = m_{A'} /2 \pi$. Meanwhile, the lower bound on the detectable DM mass is set by the energy spacing between the relevant Rydberg levels, as given explicitly in Eq.~\eqref{levelbounds}. Therefore, the mass range accessible to our superheterodyne scheme is
\begin{equation}
    5.3 \times 10^{-5}\,{\rm eV} \lesssim m_{A'} \lesssim 6.8\times10^{-4}\,{\rm eV},
\end{equation}
The above discussion concerns the highest frequency reachable in our setup. Each resonant measurement covers only a narrow bandwidth, determined by ${\rm min}[\Delta f \sim f/Q,\, \delta_s]$. Additional fine-tuning mechanisms can be incorporated into the same cavity to scan a wider range of nearby frequencies~\cite{McCulloch2024TunableResonator}. Lower-frequency ranges require larger cavities, since the cavity eigenfrequency is primarily determined by its geometry, $f_{0n0}\propto n/R$. Increasing the cavity size is therefore not a technical limitation.

As discussed above, the resonant field enhancement is $A\propto Q |\eta_{\rm field}|$. For the low-order TM$_{0n0}$ modes with $n \leq 3$, the local field-response factor $|\eta_{\rm field}|$ remains of $\mathcal{O}(1)$~\cite{sikivie1983experimental} (see App.~\ref{cavityMechanism}). Thus, $A=10^3$ can be achieved with a conservative loaded quality factor, $Q\sim10^{3}$~\cite{gohari2024ultra}, while $A=10^4$ corresponds to an optimized high-$Q$ cavity~\cite{suleymanzade2020tunable}. These benchmark values are broadly consistent with the detailed simulation results presented in App.~\ref{cavityMechanism}.

\vspace{0.15cm}

\noindent
{\bf Long-time Integration Enhanced Sensitivity:} This atomic superheterodyne approach can achieve high sensitivity through sufficiently long integration times. When the total measurement time $T$ greatly exceeds the DM coherence time $\tau_c$, the signal enters the so-called incoherent-integration regime, and the minimum detectable peak electric-field amplitude is given by~\cite{schlossberger2026fundamental}
\begin{equation}
    E_{\rm min}
    \simeq \sqrt{2}\,S\,(T\tau_c)^{-1/4},
    \label{eq:Emin_peak_main}
\end{equation}
with details given in App.~\ref{sensitivity}. Here, the electric-field sensitivity experimentally demonstrated in the latest experiment at the transition frequency $f \simeq 6.94\,\mathrm{GHz}$ is~\cite{jing2020atomic}
\begin{equation}
    S=\frac{\hbar}{\sqrt{2}\mu}
    \frac{|\tilde{P}(\delta_s)|}{|\kappa_0|}
    \overset{\min}{=}\frac{55\,\mathrm{nV}}{\mathrm{cm}\sqrt{\mathrm{Hz}}} 
    \times \left(\frac{f}{6.94\,\text{GHz}} \right)^{2/3}.
    \label{sensitivity2}
\end{equation}
where $|\tilde{P}(\delta_s)|$ is the single-sided Fourier amplitude of the optical readout at the signal frequency, $\kappa_0$ is the optimal conversion slope from the microwave-induced Rydberg energy shift to the probe transmission, and $\mu$ is the dipole matrix element of the relevant Rydberg transition. The ratio $|\tilde{P}(\delta_s)|/|\kappa_0|$ is mainly determined by the optical readout and detection chain, and is treated as a frequency-independent benchmark. The smallest measurable electric field is set by the noise floor of the superheterodyne system and the finite frequency resolution of the detection chain. The frequency dependence in Eq.~\eqref{sensitivity2} follows directly from $\mu\propto n^{\ast 2}$ and $f\propto n^{\ast -3}$ for neighboring Rydberg states, yielding $S\propto \mu^{-1}\propto f^{2/3}$.

The projected sensitivity to hidden-photon kinetic mixing is obtained by comparing the target field amplitude in \geqn{eq:Etarget} with the minimum detectable peak field in \geqn{eq:Emin_peak_main}. Assuming that the DM-induced electric field is aligned with the cavity field axis, this gives
\begin{equation}
    \epsilon_{\min} =
    \frac{\sqrt{2}\,S\,(T\tau_c)^{-1/4}}{A\,\sqrt{2\rho_{\rm DM}}} 
    \label{eq:epsmin}
\end{equation}
Using the sensitivity normalization in \geqn{sensitivity2}, with the DM coherence time $\tau_c \simeq 1/(m_{A'} v^2)$, this becomes
\begin{equation}
    \epsilon_{\min} \simeq
    7.8\times10^{-11}
    \left(\frac{m_{A'}}{1\,\mathrm{meV}}\right)^{\frac{11}{12}}
    \left(\frac{10^3}{A}\right)
    \left(\frac{T}{2\,\mathrm{hr}}\right)^{-\frac{1}{4}}.
    \label{eq:epsmin_benchmark}
\end{equation}
The scaling $\epsilon_{\min}\propto m_{A'}^{11/12}$ reflects the combination of the atomic sensitivity, scaling as $m_{A'}^{2/3}$, and the shortening of the DM coherence time, $\tau_c\propto m_{A'}^{-1}$, at higher masses.

\begin{figure}[!t]
    \centering
    \includegraphics[width=8cm]{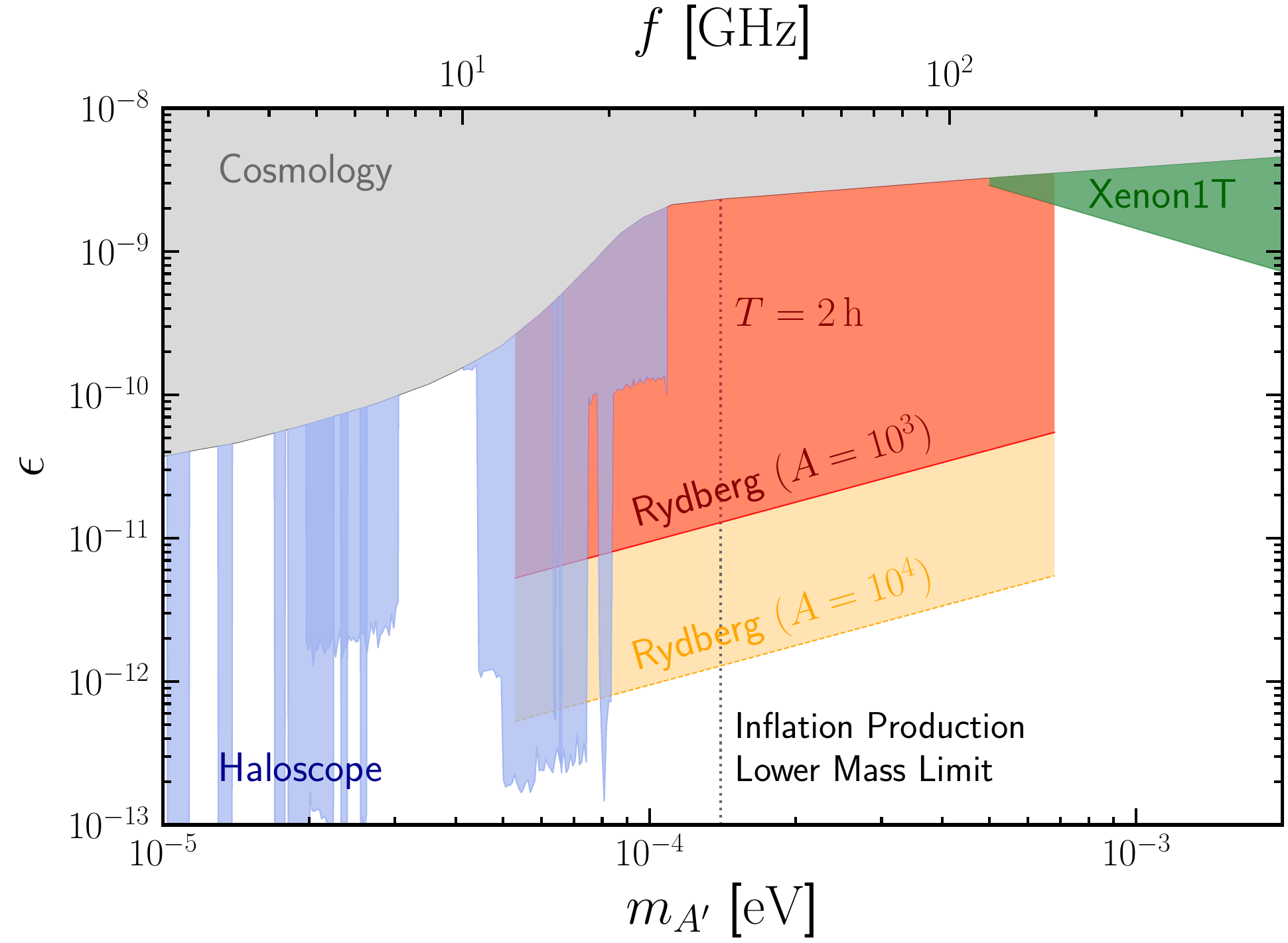}
    \caption{Projected reach of the cavity-enhanced Rydberg-atom superheterodyne search. The integration time is $2\,$hours for each tuning point. The electric-field amplification factor is set to $A=10^3$ (red) and $10^4$ (yellow). Existing bounds from haloscope searches (blue)~\cite{AxionLimits}, cosmology (gray)~\cite{Arias:2012az}, and XENON1T direct absorption (green)~\cite{XENON:2021qze} are shown for comparison. The vertical dashed line marks the lower mass limit for inflationary hidden-photon DM production\,(\ref{relic}).}
    \label{fig:limit}
\end{figure}

The resulting projected sensitivities, for integration measurement time $T=2\,{\rm h}$ per tuning point and amplification factors $A=10^3$ and $10^4$, are shown in Fig.~\ref{fig:limit}. At each tuning setting, the search remains narrowband, with width $\delta_s\sim{\rm MHz}$. By selecting different Rydberg transitions and matching the cavity dimensions to the corresponding resonant frequencies, the experiment can probe discrete transition points within this mass range.

Among the existing bounds shown in Fig.~\ref{fig:limit}, haloscope limits provide the most direct laboratory comparison~\cite{AxionLimits}, since they are likewise based on the resonant conversion of hidden photon DM into electromagnetic signals. Their sensitivity, however, deteriorates rapidly toward larger $m_{A'}$, because the resonator volume decreases with the wavelength, leaving the meV region weakly explored. The cosmological bound~\cite{Arias:2012az} arises from CMB distortions caused by the resonant conversion of hidden photons into photons in the early Universe. It therefore covers a wide mass range but is relatively weak. The XENON1T bound~\cite{XENON:2021qze,Davoudiasl:2026opo} arises from the direct absorption of solar-produced hidden photons and is most stringent in a somewhat heavier mass range.
The gray dashed line indicates the lower mass bound on vector DM implied by inflationary production, as given in Eq.\,(\ref{relic}). As shown, our proposed cavity-enhanced Rydberg superheterodyne scheme would provide the strongest limits on hidden photon DM in the mass range $8 \times 10^{-5}$--$6 \times 10^{-4}\,\mathrm{eV}$, surpassing existing cosmological bounds by $3$--$4$ orders of magnitude.

\vspace{0.35cm}

\noindent
\prlsection{Conclusion and Discussions}{.}
The detection of hidden-photon DM with masses above $10^{-4}\,\mathrm{eV}$ remains essentially unexplored. In this Letter, we propose that chip-scale Rydberg-atom superheterodyne detection, combined with cavity enhancement, provides a viable solution to this challenge. The large dipole response of Rydberg states and their GHz-scale level splittings enable sensitive detection of weak high-frequency fields, while the millimeter-scale vapor cell can be integrated into a compact high-frequency distributed cavity. With demonstrated sensitivities and integration times, together with a conservative cavity enhancement, this scheme can probe hidden-photon DM in the $5 \times 10^{-5}$--$7\times10^{-4}\,\mathrm{eV}$ mass range down to $\epsilon \lesssim 10^{-11}$. 
Because the measurement is resonant, the experiment probes narrow bandwidths around selected Rydberg transition frequencies. Fine scanning around each selected transition can be assisted by magnetic-field tuning as outlined in appendix.
In contrast to power-readout haloscope experiments, whose signal strength is limited by the cavity volume, our mechanism directly detects the DM-induced electric-field amplitude and converts the high-frequency signal into a low-frequency beat note. The same platform provides a powerful route toward precision sensing of weak, high-frequency electric fields in controlled laboratory settings.

\vspace{0.35cm}

\noindent
\prlsection{Acknowledgements}{.}
The authors thank Tsutomu T. Yanagida for discussions on the hidden-photon production mechanism and the Stueckelberg mechanism, Guanyao Huang for discussions on vapor cells, and Chaoyang Lu for discussions on magnetic-field tuning of Rydberg transitions.
B.~G.\ is supported by NSFC grant No.92565305 and Quantum Science and Technology-National Science and Technology Major Project 2021ZD0302700.
J.~S.\ is supported by the Japan Society for the Promotion of Science (JSPS) as a
part of the JSPS Postdoctoral Program (Standard) with grant number: P25018.
S.~M.\ is supported by the Grant-in-Aid for Scientific Research from the Ministry of Education, Culture, Sports, Science and Technology, Japan (MEXT), under Grant No.\ 24H00244 and 24H02244. 
S.~M. and J. S. are also supported by the World Premier International Research Center Initiative (WPI), MEXT, Japan (Kavli IPMU).
H.~D.\ is supported by cultivation Project of Shanghai Research Center for Quantum Sciences, Grant No. LZPY2024.

\begin{appendix}

\section{Cavity Enhancement of the Hidden-Photon-Induced Electric Field}
\label{cavityMechanism}

In this appendix, we clarify how the resonant cavity enhancement factor $A$ scales with the quality factor $Q$. We assume that the electric field in the cavity can be expanded in terms of a single resonant spatial mode $\mathbf{u}(\mathbf{r})$ with a time-dependent amplitude $x(t)$, namely $\mathbf{E}_{\rm cav}(\mathbf{r},t)=x(t)\,\mathbf{u}(\mathbf{r})$. Projecting Maxwell's equations onto this mode gives an effective oscillator equation,
\begin{equation}
    \ddot x + \frac{\omega_0}{Q} \dot x + \omega_0^2 x =
    \omega_0^2\,\epsilon\,E_{\rm dp}\,\eta_{\rm drive}\,\cos(\omega t).
\end{equation}
Here, $\omega_0$ is the cavity eigenfrequency, and $\omega$ is the hidden-photon frequency. The second term on the left-hand side describes damping, with energy decay rate $\omega_0/Q$, while the right-hand side is the driving term induced by the hidden-photon electric field of amplitude $\epsilon E_{\rm dp}$. The enhancement is also proportional to a dimensionless drive-overlap factor $\eta_{\rm drive}$, measuring how efficiently this drive projects onto the cavity eigenmode, as discussed below.

Solving the equation with this enhancement, the signal field sampled by atoms at position $\mathbf{r}_a$ can be written as
\begin{equation}
    E_s(\mathbf r_a,\omega) =
    \epsilon E_{\rm dp}\,
    \eta_{\rm field}(\mathbf r_a)
    \frac{\omega_0^2}{\omega_0^2-\omega^2-i\omega\omega_0/Q}.
\end{equation}
At or near exact resonance, we define the amplification factor of the induced electric field inside the cavity as
\begin{equation}
    A \equiv
    Q|\eta_{\rm field}(\mathbf r_a)|.
    \label{Afactor}
\end{equation}
The linear-in-$Q$ field enhancement above assumes that the cavity bandwidth exceeds the intrinsic DM linewidth. The DM field has a fractional bandwidth $\Delta\omega/\omega\sim v^2\sim10^{-6}$, corresponding to a coherence quality factor $Q_{\rm DM}\sim10^6$. For $Q\lesssim Q_{\rm DM}$, the resonator can respond coherently to the DM field over its linewidth, and the resonant field enhancement is well approximated by \geqn{Afactor}.

The quantity $\eta_{\rm field}({\bf r}_a) \equiv \left[\mathbf e_{\rm at}\cdot \mathbf u(\mathbf r_a)\right]\eta_{\rm drive}$ introduced above is the local electric-field response factor at the atomic position. It quantifies the cavity-field amplitude sampled by the Rydberg atoms, incorporating both the global drive--mode overlap and the local projection of the cavity mode onto the relevant atomic transition dipole direction. For a mode profile $\mathbf u(\mathbf r)$, it is defined as
\begin{equation}
    \eta_{\rm field}(\mathbf r_a) \equiv 
    \frac{
        \left[\mathbf e_{\rm at}\cdot \mathbf u(\mathbf r_a)\right]
        \displaystyle
        \int_V d^3r\,\epsilon_r(\mathbf r)\,\mathbf u^*(\mathbf r)\cdot \mathbf e_{\rm hp}
    }{
        \displaystyle
        \int_V d^3r\,\epsilon_r(\mathbf r)\,|\mathbf u(\mathbf r)|^2
    }.
\end{equation}
The Rydberg atoms measure the electric field projected along the transition-dipole direction, denoted by $\mathbf e_{\rm at}$, while the hidden-photon polarization direction is denoted by $\mathbf e_{\rm hp}$. If the atoms occupy a finite sensing volume rather than a single point, the local field should be averaged over the atomic distribution. For a sub-millimeter active atomic region placed near a field antinode of a millimeter-scale cavity, however, the pointlike approximation is usually sufficient at the order-of-magnitude level.

As an analytic example, consider an ideal cylindrical cavity with a TM$_{0n0}$ mode. The electric field is predominantly along the cylinder axis and can be written as
\begin{equation}
    \mathbf u_n(\mathbf r) =
    \hat{\mathbf z}\,J_0\left(\chi_{0n}\frac{r}{R}\right),
\end{equation}
where $R$ is the cavity radius and $\chi_{0n}$ denotes the $n$-th zero of $J_0$. If the hidden-photon polarization, the cavity electric field, and the atomic transition dipole are all perfectly aligned along $\hat{\mathbf z}$, and the atoms are placed exactly at the cavity center, $r_a=0$, then $J_0(0)=1$ and
\begin{equation}
    \eta_{{\rm field},n} =
    \frac{
        \displaystyle
        \int_V d^3r\,J_0\left(\chi_{0n}\frac{r}{R}\right)
    }{
        \displaystyle
    \int_V d^3r\,
    J_0^2\left(\chi_{0n}\frac{r}{R}\right)
    }.
\end{equation}
Using standard Bessel-function identities, this gives
\begin{equation}
    \eta_{{\rm field},n} =
    \frac{2}{\chi_{0n}J_1(\chi_{0n})}.
\end{equation}
Numerically, this factor is evaluated for the draft modes,
\begin{align}
    \chi_{01} &= 2.405,
    &
    |\eta_{{\rm field},1}| &\simeq 1.60,
    \\
    \chi_{02} &= 5.520,
    &
    |\eta_{{\rm field},2}| &\simeq 1.06,
    \\
    \chi_{03} &= 8.654,
    &
    |\eta_{{\rm field},3}| &\simeq 0.85.
\end{align}
Thus, for a local electric-field probe near the cavity center, the TM$_{020}$ and TM$_{030}$ modes can still yield an order-one field-response factor. This should be distinguished from the usual haloscope power form factor, which is smaller than the values here. If the hidden-photon polarization is not aligned with the cavity axis, the response acquires an additional projection factor of $\mathcal{O}(1)$.

\begin{figure}[!t]
    \centering
    \includegraphics[width=8cm]{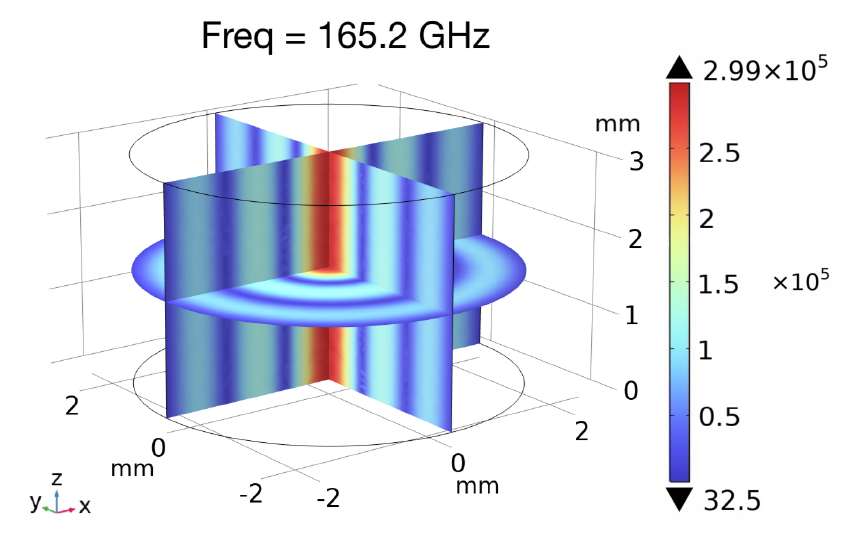}
    \caption{Simulation of the electric-field enhancement in a cylindrical cavity with radius $2.5\,$mm and height $3\,$mm. The selected cavity mode is TM$_{030}$, corresponding to a resonant frequency of $f = 165.2$\,GHz. The color scale represents the electric-field amplification factor.}
    \label{fig:cavity}
\end{figure} 

To obtain a more quantitative estimate of the cavity-induced electric-field enhancement, we performed a frequency-domain electromagnetic simulation of a cylindrical cavity by COMSOL Multiphysics 6.3. The cavity has radius $R=2.5\,\mathrm{mm}$ and length $L=3\,\mathrm{mm}$, with the side wall modeled as copper using $\sigma=5.8\times10^{7}\,\mathrm{S/m}$. The simulation was carried out at $f=165.2\,\mathrm{GHz}$, corresponding to the TM$_{030}$ mode of the cylindrical cavity. Operation at lower frequencies is expected to be easier. For an ideal closed cylinder, the resulting field-enhancement factor is very large, reaching the order of $10^5$. Because the relevant modes belong to the radial TM$_{0n0}$ family, small apertures on the two end faces are expected to produce only a minor perturbation to the field profile and quality factor, so the closed-cavity result provides a reasonable estimate of the achievable enhancement.
Taking into account additional losses introduced by the glass vapor cell, it is reasonable to assume that the atoms experience an effective field enhancement factor in the range of $10^3$–$10^4$.

A further comment concerns the comparison with the usual power scaling in haloscope experiments. The intracavity stored energy in a haloscope scales as
\begin{equation}
    U_{\rm cav}\propto |E_{\rm cav}|^2 \propto
    Q^2\epsilon^2 E_{\rm DM}^2.
\end{equation}
However, the power dissipated in or extracted from it is
\begin{equation}
    P_{\rm out} \sim
    \frac{\omega_0}{Q}U_{\rm cav} \propto
    Q\epsilon^2 E_{\rm DM}^2.
\end{equation}
Therefore, a power-readout haloscope generally has
\begin{equation}
    P_{\rm sig} \propto Q\epsilon^2,
\end{equation}
and its kinetic-mixing sensitivity improves only as
\begin{equation}
    \epsilon_{\rm min}\propto Q^{-1/2}.
\end{equation}
By contrast, if the detector directly probes the local intracavity electric-field amplitude itself, as in the Rydberg-atom scheme considered in this work, then
\begin{equation}
    E_s \propto Q \epsilon,
\end{equation}
and the corresponding sensitivity therefore scales as
\begin{equation}
    \epsilon_{\rm min} \propto Q^{-1}.
\end{equation}
The key point is that, for a coherently driven resonator, the intracavity electric-field amplitude at resonance scales linearly with the loaded quality factor, $Q$.

\section{Rydberg Superheterodyne Detection}
\label{superheterodyne}

We briefly summarize the minimal theory underlying the Rydberg-atom superheterodyne scheme used in the main text. A weak microwave signal need not produce a resolvable Autler--Townes splitting by itself. Instead, a strong local microwave field dresses the relevant Rydberg transition, converting the weak signal into a low-frequency modulation of the optical EIT transmission.

We begin with a near-resonant two-level coupling between atomic states driven by a classical field, $E(t)=E_0\cos(\omega t+\phi)$. If $\omega_0$ denotes the bare transition frequency between the two levels, the detuning is
\begin{equation}
    \Delta \equiv \omega-\omega_0,
\end{equation}
and the electric-dipole interaction defines Rabi frequency
\begin{equation}
    \Omega = d E_0,
\end{equation}
where $d$ denotes the relevant transition dipole matrix element. The detuning $\Delta$ quantifies the driving-frequency offset from resonance, while the Rabi frequency $\Omega$ characterizes the coupling strength due to the applied field.

We next consider the ladder subsystem $|1\rangle\leftrightarrow |2\rangle\leftrightarrow |3\rangle$, in which the probe and coupling fields drive the $|1\rangle\leftrightarrow |2\rangle$ and $|2\rangle\leftrightarrow |3\rangle$ transitions with Rabi frequencies $\Omega_p$ and $\Omega_c$ and detunings $\Delta_p$ and $\Delta_c$, respectively. In the rotating-wave approximation, the Hamiltonian is
\begin{align}
    H_\mathrm{EIT} =
    &
    -\hbar\Delta_p|2\rangle\langle 2|-\hbar(\Delta_p+\Delta_c)|3\rangle\langle 3| \\
    &
    +\frac{\hbar\Omega_p}{2}\bigl(|1\rangle\langle 2|+\mathrm{h.c.}\bigr) + \frac{\hbar\Omega_c}{2}\bigl(|2\rangle\langle 3|+\mathrm{h.c.}\bigr).
    \nonumber
\end{align}
A dark state $\ket{D}$ exists because one can form a coherent superposition with no admixture of the lossy intermediate state $|2\rangle$, defined by the condition $\braket{2 | H_{\rm EIT}| D} = 0$,
\begin{equation}
    |D\rangle=\frac{\Omega_c|1\rangle-\Omega_p|3\rangle}{\sqrt{|\Omega_c|^2+|\Omega_p|^2}}.
\end{equation}
When the two-photon detuning vanishes at resonance,
\begin{equation}
    \delta\equiv \Delta_p+\Delta_c=0,
\end{equation}
the system is trapped in this dark state, and the probe transmission becomes thereby maximal. In particular, for $\Delta_p=0$, the EIT maximum occurs at $\Delta_c=0$.

To implement superheterodyne detection, we introduce a strong local-oscillator microwave field, denoted by LO in the main text, tuned to the $|3\rangle\leftrightarrow |4\rangle$ transition and characterized by the Rabi frequency $\Omega_L$. The corresponding effective interaction Hamiltonian then reads
\begin{equation}
    H_L =
    \frac{\hbar\Omega_L}{2}
    \bigl(|3\rangle\langle 4|+|4\rangle\langle 3|\bigr).
\end{equation}
This coupling hybridizes $|3\rangle$ and $|4\rangle$ into dressed states
\begin{equation}
    |\pm\rangle =
    \frac{|3\rangle\pm |4\rangle}{\sqrt{2}},
    \qquad
    E_\pm =
    \pm\frac{\hbar\Omega_L}{2}.
\end{equation}
Hence, the original EIT resonance splits into two transmission peaks. For $\Delta_p=0$, the two peaks occur at
\begin{equation}
    \Delta_c =
    \pm \frac{\Omega_L}{2},
\end{equation}
so that their separation provides a direct measure of the local-field Rabi frequency of the applied microwave field,
\begin{equation}
    \Delta_c^{(+)}-\Delta_c^{(-)} =
    \Omega_L.
\end{equation}
This is the Autler--Townes splitting in EIT language.

We now introduce a weak signal microwave field on the same $|3\rangle\leftrightarrow |4\rangle$ transition, with Rabi frequency $\Omega_s$. Defining $\delta_s$ as the \emph{linear} beat frequency relative to the local oscillator, the total coupling can be written as
\begin{equation}
    \Omega_\mathrm{tot}(t)=\left|\Omega_L+\Omega_s e^{-i(2\pi \delta_s t+\phi_s)}\right|.
\end{equation}
In the small-signal regime, $\Omega_s\ll \Omega_L$, this becomes
\begin{equation}
    \Omega_\mathrm{tot}(t)\simeq \Omega_L+\Omega_s\cos(2\pi\delta_s t+\phi_s).
\end{equation}
Therefore, the two dressed-state energies acquire modulations with opposite phases induced by the signal,
\begin{equation}
    \delta E_\pm(t)=\pm\frac{\hbar\Omega_s}{2}\cos(2\pi\delta_s t+\phi_s).
\end{equation}
Physically, the weak signal does not create a new resolved doublet; rather, it periodically shifts the two Autler--Townes branches generated by the local oscillator.

The optical readout is obtained by operating on the slope of the split EIT resonance. Expanding the transmitted probe power around the operating point gives
\begin{equation}
    P(t)=P_0+\kappa\,\Omega_s\cos(2\pi\delta_s t+\phi_s),
\end{equation}
where $P_0$ is the static transmission in the presence of the local field, and $\kappa$ is the linear conversion coefficient from the microwave Rabi frequency to the optical transmission response. Since $\Omega_s=d_{34}E_s/\hbar$, the weak electric field is mapped linearly and coherently onto a low-frequency optical signal. 
Here for generality, we also keep the phase $\phi_s$.

In conclusion, ordinary ladder EIT provides a narrow optical discriminator, while the strong local microwave field converts the Rydberg transition into an Autler--Townes doublet. A weak signal field is then read out as a linear modulation of the probe transmission at the difference frequency. This is precisely the relevant regime in which extremely weak microwave or DM-induced electric fields can be detected without requiring the signal field itself to produce a spectroscopically resolved splitting.

\section{Energy Levels of Rydberg States}
\label{levels}

\subsection{Energy Eigenvalues and Transition Frequencies}

For an alkali Rydberg atom, the valence electron is nearly hydrogenic at large radii, while short-range core penetration and core polarization shift the low-$\ell$ levels from the pure Coulomb spectrum. A highly accurate starting point is therefore the Rydberg--Ritz formula~\cite{Gallagher1994RydbergAtoms}
\begin{equation}
    E_{n\ell j} =
    -\frac{R}{\left[n-\delta_{\ell j}(n)\right]^2},
    \label{eq:rydberg_ritz}
\end{equation}
where $R = 13.6\,\mathrm{eV} \simeq 3.3\times 10^{15}\,\mathrm{Hz}$ is the Rydberg energy in frequency units. Here, $n$ is the principal quantum number, $\ell$ is the orbital angular momentum, and $j$ is the total electronic angular momentum. The quantity
\begin{equation}
    n^\ast \equiv n-\delta_{\ell j}(n)
\end{equation}
is the \emph{effective principal quantum number}, and $\delta_{\ell j}(n)$ is the quantum defect. For many estimates at moderate or large $n$, it is sufficient to approximate the quantum defect as an $n$-independent constant, $\delta_{\ell j}(n)\simeq \delta_{\ell j}$.

The frequency associated with a transition connecting the two specified Rydberg levels $|n\ell j\rangle$ and $|n'\ell'j'\rangle$ is
\begin{equation}
    \nu_{n\ell j\rightarrow n'\ell'j'} =
    E_{n'\ell'j'}-E_{n\ell j}.
    \label{eq:transition_freq_def}
\end{equation}
Using the Rydberg--Ritz formula in Eq.~\eqref{eq:rydberg_ritz}, one obtains
\begin{equation}
    \nu_{n\ell j\rightarrow n'\ell'j'} \simeq
    R\left[
        \frac{1}{\left(n-\delta_{\ell j}\right)^2}
        -\frac{1}{\left(n'-\delta_{\ell' j'}\right)^2}
    \right].
    \label{eq:transition_freq}
\end{equation}
For neighboring states with similar quantum defects, the above expression can be further approximated as
\begin{equation}
    \Delta \nu \sim \frac{2 R}{(n^\ast)^3}\,\Delta n^\ast,
\end{equation}
which makes explicit the well-known $n^{-3}$ scaling of the Rydberg level spacing and thereby explains why transitions among highly excited Rydberg states naturally tend to lie in the microwave to sub-THz frequency range.

For practical estimates, it is often useful to keep a compact set of representative quantum defects. Table~\ref{tab:qd_typical} summarizes some commonly used values for Rb and Cs. Since the fine-structure dependence is small compared with the differences among orbital angular momenta, we quote a single representative value for each orbital series.

\begin{table}[t]
    \begin{ruledtabular}
    \begin{tabular}{ccccc}
        Atom & $S$ & $P$ & $D$ & $F$ \\
        \hline
        Rb & $3.13$ & $2.65$ & $1.35$ & $0.016$ \\
        Cs & $4.05$ & $3.56$ & $2.47$ & $0.04$ \\
    \end{tabular}
    \end{ruledtabular}
    \caption{Representative quantum defects for the Rydberg series of Rb and Cs. The listed values provide typical estimates for the $S_{1/2}$, $P_{3/2}$, $D_{5/2}$, and $F$ series, respectively.}
    \label{tab:qd_typical}
\end{table}

A few general trends are worth noting. First, the quantum defect decreases rapidly with increasing orbital angular momentum, as the centrifugal barrier suppresses penetration into the ionic core. Second, Rb and Cs differ substantially for low-$\ell$ states, especially in the $S$ and $P$ series. Thus, at fixed principal quantum number $n$, the Rydberg energies and intra-manifold transition frequencies of Cs can differ noticeably from those of Rb.

\subsection{Magnetic-Field Tuning of Rydberg Transitions}

A useful way to enlarge the frequency coverage of the Rydberg-atom superheterodyne scheme is to apply a static magnetic field and exploit the Zeeman splitting of the selected Rydberg levels. Without the field, the signal frequency must lie close to the bare $|3\rangle\leftrightarrow |4\rangle$ transition. The field lifts the degeneracy of the magnetic sublevels and shifts each Zeeman branch, thereby converting a discrete atomic resonance into a continuously tunable one.

For a state with total angular momentum $J$ and magnetic quantum number $m_J$, the leading magnetic-field shift is the linear Zeeman shift
\begin{equation}
\Delta E_Z = g_J \mu_B m_J B,
\end{equation}
where $g_J$ is the Land'e factor, $\mu_B$ is the Bohr magneton, and $B$ is the magnitude of the applied static magnetic field. For Rydberg states, the quadratic diamagnetic contribution can also be relevant. Taking the magnetic field along the quantization axis and using the symmetric gauge, the quadratic diamagnetic perturbation is
\begin{equation}
\Delta E_{\rm qua}
=
\frac{e^2}{8m_e}B^2\rho^2 ,
\end{equation}
where $e$ is the elementary charge, $m_e$ is the electron mass, and $\rho^2=x^2+y^2$ is the squared transverse distance of the Rydberg electron from the magnetic-field axis. For Rydberg states, $\langle \rho^2\rangle$ scales approximately as $a_0^2 n^{*4}$, where $a_0$ is the Bohr radius.

Therefore, the energy shift of state $|i\rangle$ can be written as
\begin{equation}
\Delta E_i(B)
=
g_i\mu_B m_{J,i}B
+
\frac{e^2}{8m_e}B^2
\langle i|\rho^2|i\rangle .
\end{equation}
The corresponding shift of the $|3\rangle\leftrightarrow |4\rangle$ transition frequency is then
\begin{equation}
\Delta \nu_{34}(B)
=
\Delta(gm_J)\frac{\mu_B}{2 \pi}B
+
\beta_{34}B^2 ,
\end{equation}
where
\begin{equation}
\Delta(gm_J)
\equiv g_4m_{J,4}-g_3m_{J,3},
\end{equation}
and
\begin{equation}
\beta_{34}
=
\frac{e^2}{16\pi m_e}
\left(
\langle 4|\rho^2|4\rangle
-
\langle 3|\rho^2|3\rangle
\right).
\end{equation}

For the MHz-scale frequency offsets relevant to our frequency scan, only a modest magnetic field is required. The linear Zeeman contribution gives
\begin{equation}
B
\simeq
7.1\times 10^{-5}~{\rm T}
\left(\frac{\Delta\nu_{34}}{1~{\rm MHz}}\right)
\left(\frac{\Delta(gm_J)}{1}\right)^{-1}.
\end{equation}
Thus, frequency shifts of $1$, $5$, and $10~{\rm MHz}$ correspond to fields of approximately $7.1\times10^{-5}$, $3.6\times10^{-4}$, and $7.1\times10^{-4}~{\rm T}$. In this small-field regime, the quadratic diamagnetic term is usually a correction to the linear Zeeman tuning, but it can become comparable for high-$n^*$ Rydberg states or for Zeeman branches with small $\Delta(gm_J)$. This does not introduce an additional detection channel or any new experimental ingredient; it only changes the calibration of the field-dependent transition frequency from a purely linear relation to the nonlinear form given above. A more quantitative discussion of the magnetic-field effect on frequency scanning is provided in Ref.~\cite{Chigusa:2025rqs}

For superheterodyne detection, magnetic-field tuning provides a controllable knob for shifting the selected high-frequency Rydberg transition, thereby enlarging the scan range and potentially isolating cleaner Zeeman-resolved branches. Although magnetic fields can in general distort the EIT spectrum, reduce the transmission contrast, or introduce perturbations to the microwave mode through nearby metallic or magnetic structures, these effects are expected to be mild in the weak and homogeneous-field regime considered here. A full optimization would require a combined treatment of Zeeman shifts, optical pumping, Doppler broadening, and the field dependence of the dressed-state response in a specific experimental setup.

\section{Long-Time Sensitivity Scaling for Incoherent Superheterodyne Detection}
\label{sensitivity}

When the measurement time exceeds the signal coherence time, $T>\tau_c$, the down-converted superheterodyne signal loses phase coherence during the full acquisition. Coherent accumulation of a single Fourier component is then no longer possible, and the sensitivity must instead be obtained using a power-based estimator~\cite{schlossberger2026fundamental}.

We consider an input-referred electric-field signal centered at the linear frequency $f_0$, which is written as
\begin{equation}
    E(t) =
    E_0\cos\bigl[2\pi f_0 t+\phi_0+\varphi(t)\bigr] =
    \Re\!\left[\mathcal{E}(t)e^{i2\pi f_0 t}\right],
\end{equation}
where $\phi_0$ is a constant phase offset and $\varphi(t)$ is a stochastic phase that varies over the coherence timescale $\tau_c$. The corresponding slowly varying complex envelope is then
\begin{equation}
    \mathcal{E}(t) =
    E_0 e^{i[\phi_0+\varphi(t)]}.
\end{equation}
Its single-sided power spectral density is denoted by $S_E(f)$, and the total root-mean-square electric field is
\begin{equation}
    E_{\rm rms}^2 \equiv \langle E^2(t)\rangle =
    \int_0^\infty S_E(f)\,df .
\end{equation}
Although the spectrum is centered around $f_0$, it generally has a finite distribution over frequencies $f$.

For a narrowband random signal, the spectral weight is predominantly concentrated within a finite bandwidth $B_{\rm sig}$, which is set by the inverse coherence time,
\begin{equation}
    B_{\rm sig} \sim \tau_c^{-1}.
\end{equation}
After superheterodyne conversion and linear calibration, the measured time series can be conveniently written as
\begin{equation}
    x(t) = E(t)+\xi(t),
\end{equation}
where $\xi(t)$ denotes additive white noise characterized by the amplitude sensitivity $S$, so that the corresponding one-sided noise power spectral density is given by $S^2$.

Performing a fast Fourier transform over a finite measurement duration $T$ gives a frequency resolution of
\begin{equation}
    \Delta f = \frac{1}{T}.
\end{equation}
Since $T>\tau_c$, the signal is no longer coherently confined to a single Fourier bin, but is instead spread over
\begin{equation}
    M \sim \frac{B_{\rm sig}}{\Delta f} = B_{\rm sig}T
\end{equation}
approximately independent Fourier bins. Since the phase is randomized between different coherence intervals, the complex Fourier amplitudes do not add coherently. The relevant observable is therefore the signal-band power,
\begin{equation}
    \hat Q = \sum_{k\in B} \hat S_x(f_k)\,\Delta f,
\end{equation}
where $B$ denotes the Fourier bins spanning the signal linewidth $B_{\rm sig}$, and $\hat S_x(f_k)$ is the periodogram estimator of the power spectral density at the frequency $f_k$.

The expectation value of this estimator is simply the total signal power in the band, $\langle \hat Q \rangle = E_{\rm rms}^2$. Thus, in the incoherent regime, the signal contributes through the quadratic $E_{\rm rms}^2$, rather than linearly through $E_{\rm rms}$. This is the essential difference from coherent detection.

Next, consider the noise statistics of the Fourier spectrum. In a single FFT bin, the mean noise power is
\begin{equation}
    \langle \hat S_\xi(f_k)\rangle \Delta f =
    S^2 \Delta f = \frac{S^2}{T}.
\end{equation}
For Gaussian white noise, the power in an individual periodogram bin follows an exponential distribution, so its variance is of the same order as the square of its mean:
\begin{equation}
    \mathrm{Var}\left[\hat S_\xi(f_k)\Delta f \right] \sim
    \left(S^2\Delta f \right)^2 =
    \frac{S^4}{T^2}.
\end{equation}
Summing over $M \simeq B_{\rm sig} T$ independent bins gives
\begin{equation}
    \mathrm{Var}(\hat Q_\xi) \sim
    M\frac{S^4}{T^2} \sim
    \frac{S^4 B_{\rm sig}}{T}.
\end{equation}
Therefore, the root-mean-square fluctuation of the total integrated noise power, denoted by $\sigma_Q$, is given by
\begin{equation}
    \sigma_Q \sim S^2 \sqrt{\frac{B_{\rm sig}}{T}}.
\end{equation}

The corresponding signal-to-noise ratio for the band-integrated noise-power estimator is then given by
\begin{equation}
    \mathrm{SNR} =
    \frac{E_{\rm rms}^2}{\sigma_Q} \sim
    \frac{E_{\rm rms}^2}{S^2}\sqrt{\frac{T}{B_{\rm sig}}}.
\end{equation}
Defining the minimum detectable root-mean-square field, $E_{{\rm rms},\min}$, by the condition \begin{equation}
    E_{{\rm rms},\min} = S\left(\frac{B_{\rm sig}}{T}\right)^{1/4}.
    \label{eq:Ermsmin_Bsig}
\end{equation}
Using $B_{\rm sig} \sim \tau_c^{-1}$, this can be equivalently written as
\begin{equation}
    E_{{\rm rms},\min}\simeq S\,(T\tau_c)^{-1/4}.
    \label{eq:Ermsmin_tauc}
\end{equation}
If one instead quotes the minimum detectable peak electric-field amplitude, $E_{\min} \equiv \sqrt{2}\,E_{{\rm rms},\min}$, then
\begin{equation}
    E_{\min} \simeq \sqrt{2}\,S\,(T\tau_c)^{-1/4}.
    \label{eq:Emin_peak_appendix}
\end{equation}

This result shows that, for integration times exceeding the signal coherence time, the sensitivity improves more slowly than in the coherent case. Instead of the usual $T^{-1/2}$ scaling from coherent accumulation of a monochromatic Fourier amplitude, one finds the weaker scaling law
\begin{equation}
    E_{\min} \propto T^{-1/4},
\end{equation}
which is a characteristic feature of incoherent power integration over the finite bandwidth of the signal.

Microscopically, however, $S$ is not universal: it depends on how efficiently the incident microwave electric field is converted into a measurable optical modulation. For a signal field resonant with the selected Rydberg transition $|3\rangle\leftrightarrow |4\rangle$, the electric-field amplitude is first mapped onto an effective microwave Rabi frequency as
\begin{equation}
    \Omega_s=\mu E_s,
\end{equation}
where $\mu\equiv |\langle 3|\hat d|4\rangle|$ is the relevant dipole matrix element between the two Rydberg states. Around the optimal operating point of the superheterodyne spectrum, the resulting transmission modulation is linear in $\Omega_s$,
\begin{equation}
    \delta P(t) =
    \kappa_0 \Omega_s \cos(2\pi \delta_s t+\phi_s),
\end{equation}
where $\kappa_0$ denotes the local conversion slope determined by the Autler--Townes line shape and the local-oscillator power. Thus, after calibrating back to the input electric field, the sensitivity scales parametrically as
\begin{equation}
    S = \frac{\hbar}{\sqrt{2}\mu} \frac{|\tilde{P}(\delta_s)|}{|\kappa_0|}.
\end{equation}
In this sense, the dependence of $S$ on the chosen Rydberg pair is carried primarily by $\mu$, while the remaining details of the EIT linewidth, Doppler broadening, decoherence, and local-field optimization are effectively in $\kappa_0$.

For neighboring Rydberg states, the dipole matrix element follows the well-known scaling $\mu\sim n^{*2}$. If the variation of $\kappa_0$ among nearby optimal operating points is modest, this leads to the leading-order estimate
\begin{equation}
    S \propto \mu^{-1}\sim n^{*-2}.
\end{equation}
On the other hand, the transition frequency between nearby Rydberg levels follows the Rydberg scaling
\begin{equation}
    f_{34} \sim \frac{\Delta E_{34}}{h}\sim n^{*-3}.
\end{equation}
Eliminating $n^*$ between the two scalings gives
\begin{equation}
    S \sim f_{34}^{\,2/3} \sim m_{A'}^{2/3}.
\end{equation}
This relation gives the scaling basis for the frequency dependence of the sensitivity quoted in the main text.

\end{appendix}


\providecommand{\href}[2]{#2}\begingroup\raggedright\endgroup

\end{document}